Predicting phase transition pressure in solids: a semi-classical possibility

Vladan ^elebonovi}
Institute of Physics, Pregrevica 118,11080 Zemun- Beograd, Yugoslavia

vladan@phy.bg.ac.yu
vcelebonovic@sezampro.yu

Abstract: This is a short review of the physical ideas, algorithm for calculations of the phase transition pressure, and some results of a semi-classical theory of the behavior of materials under high pressure proposed by P. Savi} and R. Ka{anin. It is based on the Coulomb interaction, supplemented by a microscopic selection rule and a set of experimentally founded postulates. The theory has been applied to cases ranging from DAC experiments to the calculation of models of planetary internal structure.

Introduction

The study of materials under high pressure (and temperature) is important in a variety of situations in astrophysics, physics and related sciences. These range from highly exotic examples such as the early Universe to laboratory experiments performed in diamond anvil cells. Determining theoretically phase diagrams of solids under high pressure is an extremely complex problem in statistical mechanics ( for example Negele and Orland,1988). It involves the choice of a proper Hamiltonian for the system under consideration, the calculation of the free energy and the determination of the regions (or points) in the parameter space in which the thermo-dynamical potentials become non-analytical functions.

The purpose of this paper is to present briefly the main physical ideas and examples of applicability of a particular semi-classical theory of the behavior of materials under high pressure. It was proposed by P. Savi} and R. Ka{anin (Savi} and Ka{anin,1962/65) and nicknamed the SK theory for short. The advantage of the SK theory is that it proposes a simple algorithm for the calculation of the phase transition pressure in solids. Apart from laboratory experiments, the SK theory has found applications in astrophysics (such as Savi},1981; ^elebonovi},1989d,1992c,1995; Savi} and ^elebonovi},1994 and references given there ).

The development of the SK theory started in 1961.,with a paper by Savi} (Savi},1961) which had the aim of exploring the origin of rotation of celestial bodies. It emerged from this work that rotation is closely related to the internal structure, and that a theory of dense matter was needed to explain it correctly. This paper has two more sections: the next one contains an outline of the basic premises of the SK theory, while the third one is devoted to a brief presentation of applications of this theory to laboratory experiments .

## Basic premises of the SK theory

The object of study in the SK theory is a mole of any material subdued to the influence of high pressure. SK is based on the idea that increasing pressure leads to excitation and ionisation of the atoms and molecules that make up the material. Translated into quantum-mechanical terms, this means that increased pressure provokes the expansion of the radial part of the electronic wave function Such an idea may seem strange at first sight, because one may be inclined to think that high external pressure leads to a "crunch". However, this problem has received a quantum mechanical treatment only about a decade ago (Ma et al.,1988), nearly three decades after the idea was used by SK.

The mean inter-particle distance $a$ is defined in the SK theory by the relation

$$N_A (2a)^3 \rho = A \qquad (1)$$

where $N_A$ is Avogadro's number, $\rho$ the mass density and A the mean atomic mass of the material. One can now define the "accumulated" energy per electron as

$$E = e^2 / a \qquad (2)$$

It can be shown (Leung,1984) that $a$, as defined above, is a multiple of the Wigner-Seitz radius.

The basic premises of the SK theory are a series of 6 statements, which limit the applicability of the theory only to first order phase transitions, and establish the ratio of the densities and accumulated energies in successive phases (^elebonovi},1989d; 1992c;1995) . Without entering into details, the main practical result is the following expression for the phase transition pressure $p_{tr}$ :

$$p_{tr} = \begin{cases} 0.5101 p_i^* ; i=1,3,5,... \\ 0.6785 p_i^* ; i=2,4,6,... \end{cases} \qquad (3)$$

where 
$$p_i^* = 1.8077 b_i (V)^{-4/3} 2^{4i/3} \quad \text{Mbar} \qquad (4)$$

$$b_i = 3(a_i^{1/3} - 1) / (1 - 1/a_i) \qquad (5)$$

and
$$a_i = \begin{cases} 6/5; i=1,3,5,... \\ 5/3; i=2,4,6,... \end{cases}$$

The symbol V denotes the molar volume of the material under standard conditions, while $i$ is an index numbering first order transitions which occur in a material. Physically realizable values of this index are determined by the selection rule

$$E_0^* + E_I = E_i^* \qquad (6)$$

where $E_I$ is the ionisation or excitation potential, and $E_0^*$ and $E_i^*$ are pressure dependent characteristic energies of the specimen, which take into account only the Coulomb part of the inter-particle interaction potential.

## Applications in laboratory experiments

The algorithm proposed by SK for the calculation of the phase transition pressure was applied to 21 different materials (^elebonovi},1989d;1992c). Apart from 19 materials, for which values of phase transition pressure were known experimentally or from various theoretical calculations ($C_6H_6; CH_4; CCl_4; C_{60}; CsI; CF_2 \equiv CF_2; CdS; Al; Ba; CaO; Kr_2; Ne; LiH; Te; TeO_2; Mg_2SiO_4; Pb; S_2; AlPO_4$) the SK algorithm was applied to hydrogen and helium. They were included in the calculation because of their astrophysical importance, although the actual existence of phase transitions under high pressure in them has not yet been experimentally confirmed (Narayana et al.,1998).

The relative discrepancies between the values of the phase transition pressure calculated within SK and those existing in the literature vary between nearly 0 and 30 % (^elebonovi},1992c). These differences are due to a variety of factors.

For example, the precision of experimental data and of the input parameters in the calculations, accounts for approximately $\pm 10$ %. An important source of the discrepancies is the form of the inter-particle interaction potential. The SK takes into account only the pure Coulomb part of the interaction potential, without the contributions of the charge distribution overlap, and of the dispersive and repulsive forces. Now, the relative contribution of the dispersive and repulsive forces to the full inter-particle potential is minimal for: C,H,N,O. Interestingly, the discrepancies between the SK and experimental values of the phase transition pressure is also minimal for the hydrocarbons (^elebonovi}, 1992c).

Taking into account the existence of the charge distribution overlap gives rise to three additional terms in the expression for the accumulated energy .The sum of these terms is pressure dependent, and it can be positive, negative or zero. The existence of these terms induces an error in the calculated values of the phase transition pressure, and the magnitude of this error is also pressure dependent. Due to space limitations, we have here outlined the influence of only two factors which contribute to the relative discrepancies between the SK values of the phase transition pressure, and those obtained in experiments. A detailed account is avaliable in (^elebonovi},1992c).

Instead of a conclusion, several comments concerning SK are in order . The big advantage of this theory is its physical and mathematical simplicity. On the other hand, this simplicity necessarily induces discrepancies between the experimental values of the phase transition pressure and those calculated within this theory. Pushing the reasoning further, the existence of these discrepancies opens up the possibilites  for  improving the basic assumptions of  SK, by rendering them more complex. Some preparatory work in this direction has already started.